\documentclass[runningheads]{svmult}

\usepackage{makeidx}   
\usepackage{graphicx}  
\usepackage{subeqnar}  
\usepackage{multicol}  
\usepackage{physprbb}  
\makeindex             

\begin{document}
\title*{Astrophysical Applications of Tunable Imaging Filters
for the VLT}
\toctitle{Astrophysical Applications of Tunable Imaging
\protect\newline Filters for the VLT}
%
%
\titlerunning{Applications of Tunable Imaging Filters for the VLT}
%
\author{Heath Jones}
\authorrunning{Heath Jones}
%
%
\institute{European Southern Observatory Chile, Casilla 19001, Santiago, Chile}

\maketitle              

\begin{abstract}
Tunable imaging filters have been used for a variety of science programmes
on the Anglo-Australian and William Herschel Telescopes during the last 
five years.
This contribution describes these novel devices and reviews the science (both
Galactic and extragalactic) done with them. 
Possible strategies for implementing a tunable filter at the VLT are also
discussed. Significant scientific potential exists for a tunable filter 
on the VLT, particularly in the years before such capability becomes
available on 8 -- 10~m-class telescopes elsewhere.
\end{abstract}

\section{Introduction}

The Taurus Tunable Filter (TTF) instruments \cite{blandjon98a}\cite{blandjon98b}
at the Anglo-Australian (AAT) and William Herschel Telescopes (WHT)
have seen use in many key areas of astrophysics. Low-redshift science has included 
studies of brown dwarf atmospheric variability and the identification of 
optical counterparts to Galactic X-ray sources. At high-redshifts,
science has been driven by measurement of the cosmic star-formation history, 
identification of galaxy clustering around high-redshift QSOs, deep imaging 
of jet-cloud interactions in powerful radio galaxies, and the detection 
of a large ionized nebula around a nearby QSO. 

This paper describes the characteristics of these instruments 
and the future role they could play at the VLT.


\section{Tunable Filters}

\subsection{The Taurus Tunable Filter (TTF)}

A tunable filter \cite{atherton81}
is a special type of Fabry-Perot interferometer incorporating 
three features that traditional astronomical Fabry-Perot instruments 
lack.\footnote{Note that alternative technologies exist for tunable imaging, 
although none have been yet been found applications for night-time 
astronomy \cite{bland00}.
}
A tunable filter:
(1) can move its parallel glass plates over a large range, (2) has
anti-reflection coatings optimised over a broad range of wavelengths, and,
(3) operates at much narrower plate spacings than traditional devices.
These characteristics mean that tunable filters operate at lower
resolving powers (${\cal R} = 100$ to 1000) than traditional instruments.

The first Taurus Tunable Filter (TTF; \cite{blandjon98a}\cite{blandjon98b})
was introduced at the Anglo-Australian Observatory (AAO) in early 1996 
by J.~Bland-Hawthorn. This red device operates in the range 6500 --
9500 \AA\ at the aforementioned resolving powers, thereby giving an adjustable
passband width of 6 to 65 \AA. Two years later a second Fabry-Perot 
coated for 3700 -- 6500 \AA\ gave the potential for tunable imaging across
the full optical range. 
Since the Fabry-Perot is an interference device, many orders of interference
are present simultaneously. Therefore, one needs to use a blocking filter to 
remove light from all but the one order of interest. 
There are a dozen different $\sim 200$ -- 300~\AA-wide
blocking filters used with TTF. More details on the TTF instruments
can be found at the AAO's TTF Home Page (http://www.aao.gov.au/ttf/).

\begin{figure}[t]
\begin{center}
\includegraphics[height=.45\textheight, angle=270]{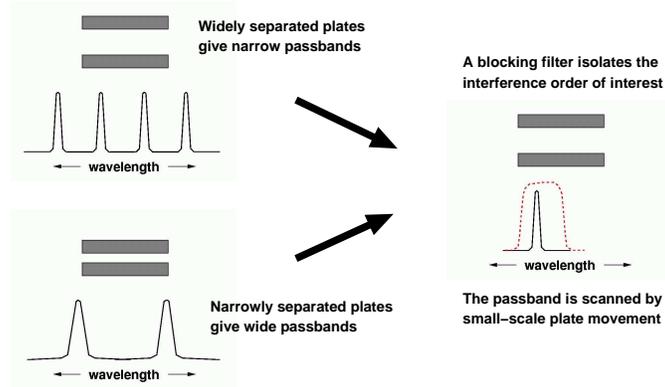}
\end{center}
\caption[]{Tuning the filter to the desired passband width and wavelength.}
\label{tuning}
\end{figure}

TTF is used in a cassegrain-mounted focal reducer (Taurus-2), which consists
of a simple camera-collimator arrangement with a straight-through
optical path. The optical train of Taurus-2 has aperture and focal plane 
filter wheels at the telescope focus, then the collimator, 
two further wheels in the parallel beam for the Fabry-Perot and 
pupil plane masks, the camera and finally the detector. The design of Taurus-2 
is very similar to that of the FORS instruments on the VLT. 
There are identical copies of Taurus-2 at both the AAT and WHT, although
Taurus-2 is no longer offered at the WHT. Taurus-2 at the AAT has
been scheduled with tunable filters in continuous semesters since 1996. 

Tuning is achieved through controlled changes to the separation between
the glass plates: initial wavelength and bandpass selection is made
by making a large adjustment; subsequent scanning is done through much
smaller changes (Fig.~\ref{tuning}). 
The plates are moved and stabilised by electronics
attached to the outside of the focal reducer.

\subsection{CCD Charge-Shuffling and Tuning}

\begin{figure}[t]
\begin{center}
\includegraphics[height=.6\textheight,angle=270]{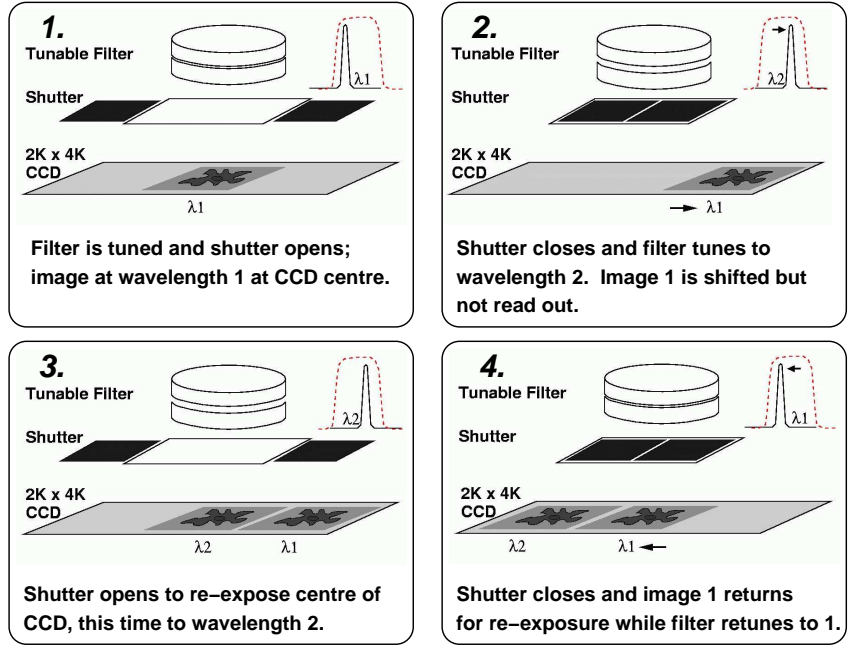}
\end{center}
\vspace{-5mm}
\caption[]{Synchronising charge-shuffling with tuning of the filter.}
\label{chargeshuff}
\end{figure}

A further development at the AAO was the synchrony of filter tuning with
the shuffling of charge on the CCD. The basis of the technique is the use
of two or more {\sl different} regions of the {\sl one} CCD frame
to image the sky at different wavelengths. 
Exposure of a particular CCD region/wavelength combination
can be made many times before the CCD frame is finally read-out.
Figure~\ref{chargeshuff} illustrates the technique. By splicing
the multiple exposures of each band, variable conditions during the total
imaging time are effectively averaged out. Hence, precise differential photometry
is possible in conditions when it otherwise would not.
Of course, a portion of the CCD frame must be sacrificed to allow extra room
for the shuffle. However, the coincidental introduction 
of over-sized $2 \times 4$~K CCDs at the AAT circumvented this problem.

\section{Science from Tunable Filters on the AAT and WHT}

Many areas of astrophysics have utilised the tunable filter
to undertake a diverse range of science. These projects point the way toward
the science possible with a tunable filter on an 8~m-class telescope.
The following is a representative (rather than comprehensive) list of 
recent results with the Taurus Tunable Filter (TTF) on the two 
4-m telescopes.

\begin{figure}[t]
\begin{center}
\includegraphics[height=.57\textheight, angle=270]{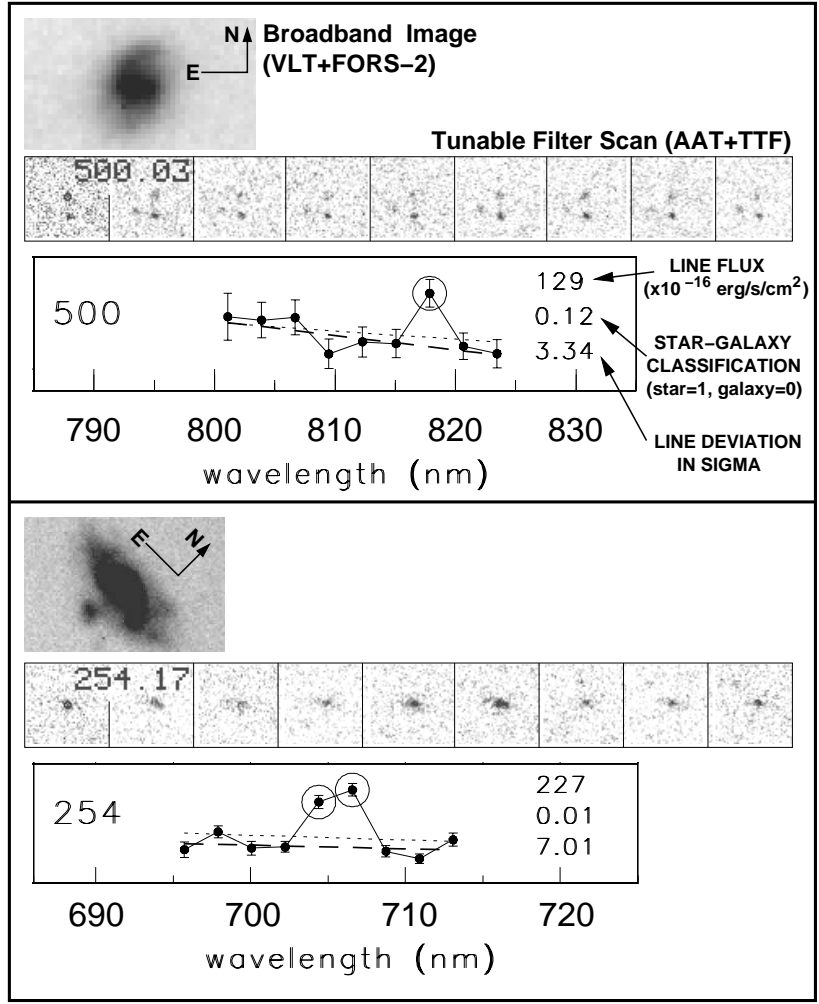}
\end{center}
\vspace{-5mm}
\caption[]{Example emission-line galaxies found with the TTF for \cite{jones01b}.}
\label{exampleGalaxies}
\end{figure}

\begin{itemize}
\item {\bf Extended Nebula Around a Nearby QSO.}~~Shopbell, Veilleux and
Bland-Hawthorn \cite{shopbell99} obtained deep H$\alpha$ imaging of the
$z = 0.0638$ quasar MR~2251$-$178. This revealed ionised gas over a
200~kpc region around the quasar, suggesting its photoionisation of 
a surrounding HI gas envelope as the cause of the emission, rather than a
merger event or interaction.
\item {\bf Field Populations of Star-Forming Galaxies.}~~The scanning
ability of TTF was used by Jones and Bland-Hawthorn \cite{jones01b}
to search for the emission-line signatures of star-forming galaxies
at $z > 0.1$. This narrowband selection found excess numbers of line-emitters
over those from traditional broadband-selected redshift surveys, 
implying higher star-formation densities over these redshifts.
Figure~\ref{exampleGalaxies} shows two TTF-selected galaxies from this survey.
\item {\bf Searching for Weather in Brown Dwarfs.}~~Tinney and Tolley
\cite{tinney99} used charge-shuffled time-series imaging of brown dwarfs
in two passbands sensitive to variations in effective temperature.
Variability was found in one of the two stars surveyed, over observations 
spanning one-third of its rotation period, indicative of surface features.
\item {\bf High-Redshift Gravitationally Lensed Galaxies.}~~Hewett and
collaborators \cite{hewett00} 
have used the tunable filter to search for gravitationally
lensed galaxies at $z \sim 3$. Giant, bulge-dominated $z \sim 0.4$ ellipticals
are identified as potential lenses through anomalous emission-lines
in 2dF Galaxy Redshift Survey spectra. TTF is tuned to the line to see
if a $z \sim 3$ galaxy (being lensed by the elliptical) is responsible for
the emission.
\item {\bf Time Series Photometry of a Stellar X-Ray Source.}~~Time-series
photometry of the X-ray star V2116 Ophiuchi was undertaken by Deutsch, Margon
and Bland-Hawthorn \cite{deutsch98}. The tunable filter was tuned to
the prominent O{\sc i} $\lambda8446$ line in this object, thought
to pulse in-phase with the X-ray source. These observations ruled-out any 
such variability.
\item {\bf Galaxy Clustering Around High-Redshift QSOs.}~~Baker and
collaborators \cite{baker01} tuned TTF to the $z=0.9$ quasar
MRC~B0450$-$221 to search for nearby clustering galaxies. Nine galaxies
were found with emission-lines matching [O{\sc ii}] at the redshift of the
QSO. Of the five accessible for spectroscopic follow-up,
three were positively identified with [O{\sc ii}] and another with a possible
line detection.
\item {\bf Warm Ionised Gas Around Nearby Radio Galaxies.}~~TTF was
used on the WHT by Tadhunter and collaborators \cite{tadhunter00} for
H$\alpha$ imaging of two radio-galaxies at $z = 0.24$ and 0.09. Faint
emission-line structures were found beyond the radio axes,
in addition to the usual bright structures along the radio jets.
\end{itemize}
Other tunable filter projects currently in progress include imaging
of filamentary structures in edge-on spiral galaxies, star-formation regions in
nearby elliptical galaxies, H$\beta$ imaging in face-on spirals and
H$\alpha$ imaging of nearby galaxy cluster cooling flows. It is likely that such
scientific diversity would continue for a tunable filter at the VLT.

\section{A Tunable Filter at the VLT}

There is currently no tunable filter capability on VLT nor
on any 8 -- 10~m-class telescopes elsewhere. However, the OSIRIS instrument
\cite{cepa00} planned for the 10~m GranTeCan telescope will have tunable
imaging, as will the SOAR telescope currently under construction in Chile
\cite{cecil00}.

Tunable filters are commercially available from Queensgate Instruments in 
several sizes, along with the associated control electronics. While a detailed
technical assessment is yet to be done, an informal study has shown
that a 116~mm tunable filter can be accommodated in FORS-2 if the
upper of the two grism wheels were removed \cite{jones01a}. The full 
spectroscopic capability of FORS-2 is preserved, although more frequent grism
changes are required. In such a case, the echelle mode would be lost from 
FORS-2 but could be incorporated into FORS-1 instead. Most importantly,
no major hardware modifications are required and the control electronics come 
from Queensgate. Software would need to be upgraded to include control of 
the instrument, to maximise observing and calibration efficiency. More 
discussion of the various options available can be found in \cite{jones01a}. 

A tunable filter on the VLT would have the ability to tune {\sl between}
the brightest OH night-sky lines. Moreover, the narrow bandpass 
would increase the usefulness of FORS during bright-time. 
As a 3-D survey instrument, a tunable filter would complement the multi-object
spectroscopic capabilities of the FORS instruments well. Tunable filters
avoid the problems of sky subtraction experienced with other 3-D
devices such as IFUs.
Furthermore, none of the special reduction techniques used for kinematic 
Fabry-Perot data are required for tunable filter data because of the much 
lower spectral resolutions utilised.


\section{Summary}

Tunable filters have been used for many extragalactic and Galactic
stellar programmes at the AAT and WHT over recent years. The ability
to synchronise charge-shuffling with tuning of the filter has allowed
precise differential imaging to be undertaken by some of these programmes,
even in variable conditions. Implementing a tunable filter at the
VLT in FORS appears technically feasible. The primary operational change 
would be more frequent grism installations to maintain all the
spectroscopic modes currently available.

%

\end{document}